\documentclass[aps,prl,10pt,twocolumn,groupaddress,floatfix]{revtex4-2}
\usepackage{amsmath, amsthm, amssymb, amsfonts}
\usepackage{natbib}
\usepackage{graphicx} 
\usepackage{hyperref}
\usepackage{color,xcolor}
\usepackage{dcolumn}
\usepackage{siunitx}

\definecolor{orange}{rgb}{0.99,0.5,0.31}

\newlength{\lengthofminus}
\newlength{\lengthofone}
\begin{document}

\title{Micro-Electro-Mechanical System Vapor Cells With Passivated Internal Cavities}

\date{\today}

\author{Rajesh Pandiyan}
\author{Sanyasi Bobbara}
\author{Somayeh Mirzaee}
\author{Su-Peng Yu}
\author{Ruoxi Wang}
\author{Adam Sibenik}
\author{Reza Kohandani}
\author{James P. Shaffer}
\email{e-mail: jshaffer@qvil.ca}
\affiliation{ Quantum Valley Ideas Laboratories, 485 Wes Graham Way, Waterloo, Ontario N2L 6R1, Canada}

\begin{abstract}
Micro-Electro-Mechanical, so called 'MEMs,' vapor cells are a key component in atom-based quantum sensors, such as clocks, gyroscopes, electric field sensors and magnetometers. MEMs vapor cell fabrication for Rydberg atom radio frequency sensors is particularly demanding. The Rydberg states used for the sensor can shift in a constant electric field which can be generated by the internal surfaces of the vapor cell cavity. The ratio of the detection wavelength to vapor cell size can span a large range, meaning that the radio frequency field- vapor cell interaction is a critical design consideration. In many radio frequency sensing cases, there is a desire to minimize the interaction between the vapor cell and the target radio frequency field, as well as assure that every vapor cell behaves uniformly. These criterion favor MEMs vapor cells with low background electric fields. Known inert, organic coatings cannot survive the bonding temperatures required for conventional anodic bonding of a MEMs vapor cell. Applying inert, organic coatings to the internal cavities of MEMs vapor cells is a longstanding challenge. In this paper, we present a low temperature bonding scheme that is compatible with coating the internal cavity of a MEMs vapor cell with Octadecyltrichlorosilane (CH$_3\,$(CH$_2$)$_{17}\,$SiCl$_3$, OTS). The coating prevents the Cs used in the vapor cell from sticking to the walls. Spectral linewidths of $\sim300\,$kHz are obtained using Rydberg spectroscopy, with energy shifts corresponding to electric fields $<$10$\,$mV$\,$cm$^{-1}$. 
\end{abstract}
\maketitle

One key component of any Rydberg atom-based RF sensor is the vapor cell that contains the atoms \cite{Sedlacek2012,Fan2015}. Microfabricated, i.e. MEMs, vapor cells \cite{Kitching04}, are important for quantum technologies like magnetometers, clocks, gyroscopes and radio-frequency sensors \cite{Schwindt04,Knappe04,Douahi07,Gyroscoperev,Sedlacek2012,Hadi2020,MEMSrev}. In contrast to glass-blown vapor cells, MEMS vapor cells are highly uniform and can be engineered more easily for specific applications. Vapor cell design and fabrication for Rydberg atom-based RF sensors are particularly complicated, because Rydberg atoms are uniquely sensitive to electric fields and collisions. The size of the vapor cell is comparable to the wavelengths of RF electromagnetic fields so it can easily transition through the Rayleigh to Mie to optical scattering regimes \cite{Hulst57}. Line broadening, frequency shifts, RF scattering, unintended alkali metal coatings of the vapor cell walls, and manufacturing difficulties (bonding, outgassing, ….) associated with the vapor cell plague efforts to capitalize on the unique advantages Rydberg atom-based RF sensors offer. 

Conventional MEMs and glass-blown vapor cells for RF sensing are known to generate electric fields that cause shifts and broadening of the spectral lines that reduce sensitivity and accuracy. We have shown in prior experiments that alkali atoms bind to oxygen on a surface, polarize and generate these fields \cite{Sedlacek16}. Most surfaces in MEMs vapor cells are oxides, including those of Si, which have a native oxide layer unless specially prepared. Other effects are desorption at the sites where the lasers pass through the vapor cell windows \cite{Holloway2025}. Background gasses remaining in the vapor cell can also broaden the spectral lines. The bonding process can generate background gas, but these gasses can be pumped away using getters after the vapor cell is filled or are reacted by the alkali to form solids, i.e., alkali gettering. Electric field mitigation is of central importance.

In this paper we use unique low temperature bonding and surface preparation methods to coat and passivate the internal cavity of a Cs vapor cell with Octadecyltrichlorosilane (CH$_3\,$(CH$_2$)$_{17}\,$SiCl$_3$, OTS). While the SiCl end of the molecule binds to the surface, the methyl group exposed to the interior of the vapor cell has filled orbitals. A self-assembled monolayer (SAM) is formed on the surface that is inert, preventing Cesium from bonding to it. Cross-linking of the OTS molecules prevents Cesium from finding its way to the glass and silicon surfaces (the silicon surface is usually a native oxide surface). The OTS coated vapor cells, while critical to Rydberg atom RF sensors, are also of use for work in clocks, gyroscopes and magnetometers because the OTS coatings can also reduce spin depolarization when the atoms collide with the walls of the vapor cell, increasing the coherence time of the optically prepared atoms \cite{Straessle2014}. We obtained electromagnetically induced absorption (EIA) spectral linewidths as narrow as $300\,$kHz in the OTS coated MEMs vapor cells. The line shifts were correspondingly low. We estimate the background electric field to be $< 10\,$mV$\,$cm$^{-1}$ derived from the polarizability of the Cs atom in the $42\textrm{P}_{3/2}$ state. Testing was accomplished using a novel, Doppler free, collinear excitation scheme for Cesium \cite{Shaffer2018,Bohaichuk2023}.

\begin{figure}[!h]
   \centering
   \includegraphics[width=0.5\textwidth]{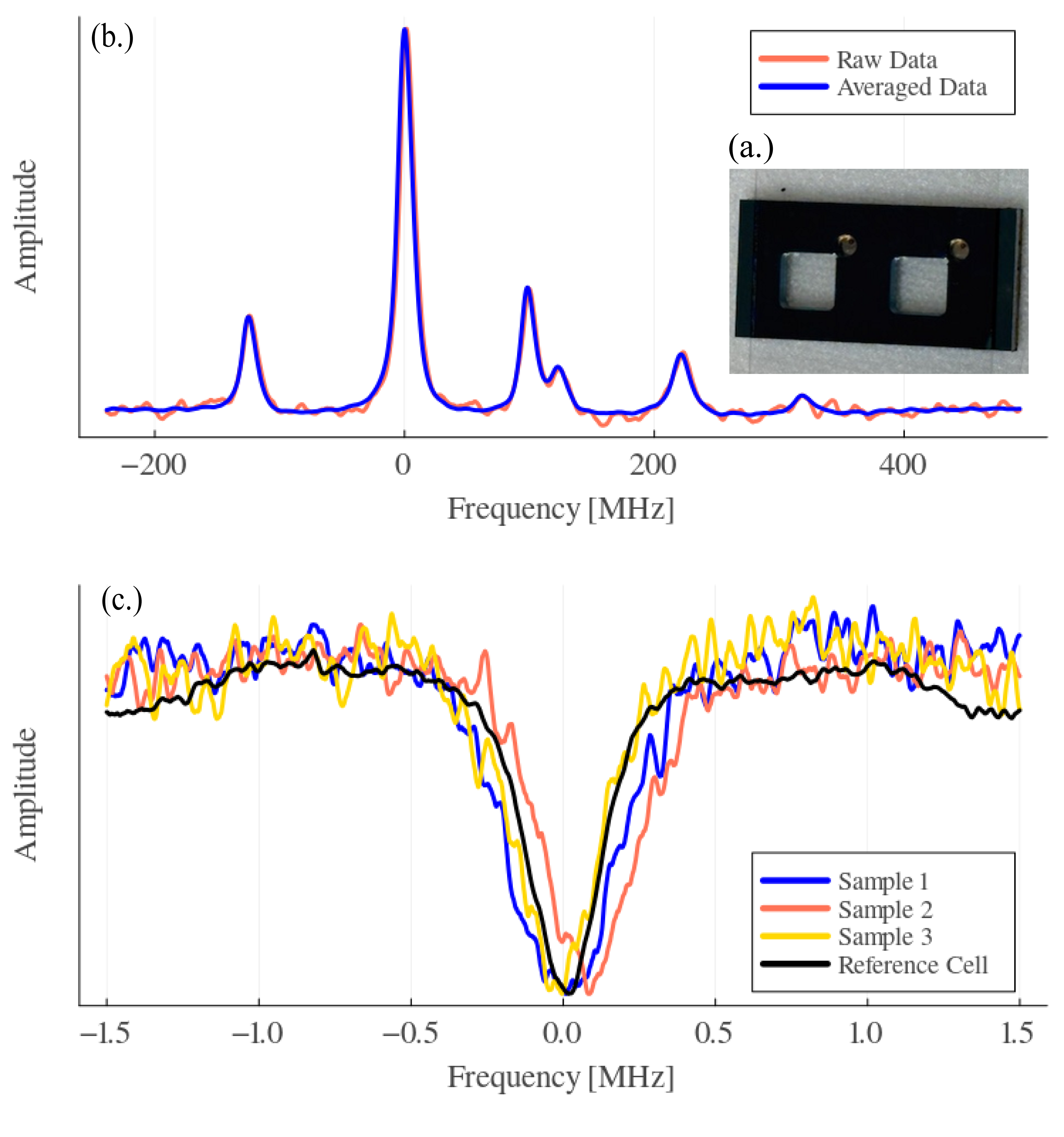}
   \caption{(a.) OTS coated vapor cell. (b.) The figure shows a single shot and averaged saturated absorption trace from an OTS coated vapor cell. (c.) The figure shows the spectral lineshapes from the three MEMs vapor cell samples used for the experiments in comparison with the data obtained from the reference vapor cell. The EIA spectra have spectral linewidths in the vapor cells of $2\pi \times 390\,$kHz (sample 1), $2\pi \times 330\,$kHz (sample 2), and $2\pi \times 290\,$kHz (sample 3). The EIA spectral linewidth in the reference vapor cells was $2\pi \times 290\,$kHz. The line shifts were measured to be $2\pi \times -5\,$kHz (sample 1), $2\pi \times 83\,$kHz (sample 2) and $2\pi \times 19\,$kHz (sample 3) relative to the line center of the reference vapor cell. The Rabi frequencies of the lasers were $2\pi \times 1\,$ MHz ($895\,$nm), $2\pi \times 1.5\,$MHz ($636\,$nm) and $2\pi \times 160\,$kHz ($2262\,$nm). The respective laser spectral linewidths were $2\pi \times 600\,$Hz, $2\pi \times 2\,$kHz and $2\pi \times 20\,$kHz. The transit time broadening was estimated to be $2\pi \times 240\,$kHz. The overlap region of the laser beams was $400\,\mu$m in diameter. }    \label{fig:spectra}
\end{figure}

The vapor cell consists of a Si frame and two borosilicate glass windows. Other configurations and numbers of bonding layers are possible. Examples of other types of vapor cells can be found in Refs. \cite{Hadi2020,Jaime2022,amarloo2024}.  One window is bonded to a Si frame to make a so-called ‘base frame,’ in air. The base frame and the second window are coated with an OTS SAM in an inert environment. The base frame is filled with a small amount, 15-25 nL, of pure Cs using piezo-electric dispensing. Chemical processes for filling the vapor cell with Cs are also possible \cite{JKitching2004}. After the base frame is charged with Cs, the base frame and second window are placed in a bonding apparatus. The bonding chamber is subsequently pumped down to $<10^{-6}\,$Torr. After the system reaches the desired pressure, the window and base frame are contacted and bonded at $\sim 140^\circ\,$C.

The OTS coating method is a wet chemical adsorption process where the plasma treated components are immersed in a solution of OTS molecules to form an OTS SAM \cite{Straessle2014, chi2020advances, chi2021comprehensive, yi2008method}. Optimization of the surface layers, specifically chemical polishing and thermal SiO$_2$ growth on the Si cavity walls, is necessary to form the OTS SAM. OTS forms the SAM by reacting with hydroxyl (-OH) groups on the surface. Without a smooth SiO$_2$ layer, OTS would not properly adhere, resulting in poor coverage and weak bonding, compromising the properties of the coating \cite{wang2003growth, onclin2005engineering}. The interfacial layer, specifically SiO$_2$, also enhances the glass//SiO$_2$/Si bond strength while allowing the process to occur at lower temperatures, $<200^\circ$C. The SiO$_2$ layer acts as a mediator, facilitating the adhesion. Other metal oxide layers, such as Al$_2$O$_3$ \cite{mitchon2006growth} and TiO$_2$ \cite{pratiwi2020self}, will behave similarly.

Commercially available intrinsic, double side polished, float zone Si:$\langle$100$\rangle$ wafers, having a high resistivity, $>20\,\mathrm{k}\Omega\,$cm, are used. The Si wafer is a standard diameter of $100\,$mm with $1.5\,$mm thickness. The RMS surface roughness is no greater than $0.15\,$nm. Double side polished borosilicate glass is obtained from Schott for the windows. The glass wafer also is a standard $100\,$mm diameter with a thickness of $500\,\mu$m. The surface roughness is less than $0.2\,$nm. The wafers are cleaned and rinsed with de-ionized water (DI) to remove any particles. To remove other organic residues and contaminants, the Si wafers are cleaned and hydrophilized in RCA (Standard Clean -I) solution (5:1:1 mixture of H$_2$O:NH$_4$OH:H$_2$O$_2$) and (Standard Clean -II) solution (5:1:1 mixture of H$_2$O:HCl:H$_2$O$_2$) at a temperature of $75-80^\circ \,$C for $15\,$min, followed by a DI rinse before being dried with pure Nitrogen gas (N$_2$).

Laser machining is used to make the channels, cavities and side pockets in the Si wafer.  Each test chip consists of a $10\,\textrm{mm} \times 20\,\textrm{mm}$ section of the wafer comprising two vapor cells, FIG.~\ref{fig:spectra}a. Each cavity is $4\textrm{mm} \times 4\textrm{mm}$ and is fluidly coupled via a $300\,\mu\textrm{m} \times 300\mu\textrm{m}$ channel of depth $750\mu\textrm{m}$ to a $1\textrm{mm}$ diameter hole used to fill the Cs. After laser micromachining, the Si wafer is cleaned using ultrasonication with Acetone, Methanol, Isopropanol, and DI, in series. After the ultrasonic cycle, the wafer is cleaned and hydrophilized in piranha solution (a 4:1 mixture of H$_2$SO$_4$:H$_2$O$_2$) for $10\,$mins, followed by DI rinse and N$_2$ drying to remove remaining debris. 

Wet etching is used to smooth the vertical sidewalls of the Si frame. Several methods have been tested to smooth the side walls. Anisotropic wet etching by KOH (45$\%$) can be employed. A $10\,$min etch at $75^\circ\,$C gives satisfactory results, resulting in clean, smooth surfaces where the crystallographic planes are observed with scanning electron microscopy. For this work, we implemented an isotropic etching recipe using HNA \cite{gad2001mems}. The HNA composition of CH$_3$COOH:HNO$_3$:HF ($10:80:10$) with an etching time of $7\,$mins, followed by a DI rinse and drying by N$_2$ produced better results. 

Next, a high quality $150\,$nm layer of low stress, low surface roughness SiO$_2$ is grown on the Si wafer. The coating is produced by dry oxidation at $\sim 1000^\circ$C. The glass and Si wafers are diced into the $10\,\textrm{mm} \times 20\,\textrm{mm}$ subsections for making the vapor cells. After dicing, the SiO$_2$ coated silicon chips and the glass chips are again cleaned ultrasonically. 

Sequential plasma treatment, using a combination of Oxygen and Nitrogen plasmas, is applied to the wafers to activate the surfaces for bonding (YES-Asher). The initial Oxygen plasma treatment is used to clean the surface and introduce reactive Oxygen species, i.e. Hydroxyl groups, which form a thin oxide layer. A Nitrogen plasma treatment further modifies the surface by introducing Nitrogen-containing functional groups. The sequential process creates a highly reactive and hydrophilic surface, which is beneficial for strong adhesion, $> 10\,$MPa, verified through pull testing. Both plasma treatments were applied for $60\,$s with an RF power of $400\,$W. The chamber pressure is maintained at $\sim 360\,$mTorr. Oxygen and N$_2$ gas are introduced into the plasma chamber at a rate of $\sim 90\,$sccm and $\sim 90\,$sccm, respectively. Since the surfaces are highly hydrophilic, no further hydroxylation of the surfaces is required.

The glass and Si chips are brought into contact with each other at room temperature by placing the activated surfaces together. The bond strength is weak because of the low bond energy of Si-OH which forms Van der Waals and hydrogen bonds. Increased bonding strength is achieved by applying voltage in an anodic bonding-like process. The resulting electrostatic interactions and chemical affinities facilitate the formation of covalent bonds consisting of Siloxane bonds, but more critically, the bonding of Silicon Oxynitrides at the interface. When the temperature is stabilized at $380^\circ$C, a DC voltage of $500-800\,$V is applied to the electrodes. The current between the electrodes is measured until the current, $\sim 1\,$mA falls to a constant value, $\sim 0.005\,$mA.

The OTS SAMs are prepared from a solution of OTS ($95\%$) and n-Hexane ($99\%$). The coating is applied in an inert environment to prevent OTS oxidation. We tried different concentrations of OTS and found that a concentration of $7.61 \times 10^{-3}$M worked best. The prepared solution required a maturation time of $90\,$minutes to stabilize the components and promote appropriate ordering of the SAM. The base frame and the capping glass are immmersed in the solution for two $15\,$minute cycles followed by a $30\,$minute cycle. The capping window is processed through the same cycle as the base frame. After each cycle, the part is washed with Hexane and dried with N$_2$. After the final rinse, the parts are placed on a hot plate and baked at $110^\circ$C for $30\,$minutes, transferred to a vacuum chamber at $<10^{-6}$Torr, and baked at $120^\circ$C overnight for outgassing.

After the baking process, a small quantity of pure Cs, $\sim 15-25\,$nL, is piezo-electrically dispensed into the side pocket of the OTS coated base frame. The capping window is bonded to the base frame at low temperature enabled by the ability of the bonding method described here to operate at temperatures, T$\sim 140^\circ$C. The low temperature bonding reduces outgassing during the bonding process and can preserve the OTS SAM. The Cs filled base frame assembly and the capping window are loaded into the bonding chamber. The chamber is evacuated to $\sim 10^{-6}$Torr. A manipulator is lowered to make a tight pre-contact bond between the base frame and the capping glass. The temperature of the stack is gradually raised to $140^\circ$C, in controlled steps to prevent thermal stress or damage to the OTS coating. A voltage of $-1500 - -1700\,$V is applied, initiating the final bonding step, which lasts for $90-120\,$minutes. The bonding process is completed when the current, $\sim 0.06\,$mA, decays to a residual value of $\sim 0.002\,$mA. The bonded chips are allowed to cool naturally to room temperature, then removed from the chamber for testing. 

The SAMs were characterized by atomic force microscopy (AFM) and contact angle (CA) measurements. The CA of a water drop and a liquid cesium drop on the OTS coating is an indicator of SAM quality; a large CA indicates a more hydrophobic and hence closer-packed film. OTS SAM deposition was confirmed by water CA measurements, which indicated a contact angle of $<5^\circ$ when measured on bare RCA-cleaned and plasma activated hydroxylated Si and glass substrates. The CA of the OTS SAM is $\ge 100^\circ$ for both the water\cite{jung2009characterization}  and Cs droplets. AFM was used to confirm the uniformity of the SAM surface on the microscopic scale. The RMS surface roughness was measured to be between $1.7-2.2\,$nm and $0.6-0.9\,$nm, on the SiO$_2$ coated Si sidewalls and glass, respectively.

\begin{figure*}[t]
   \centering
   \includegraphics[width=\textwidth]{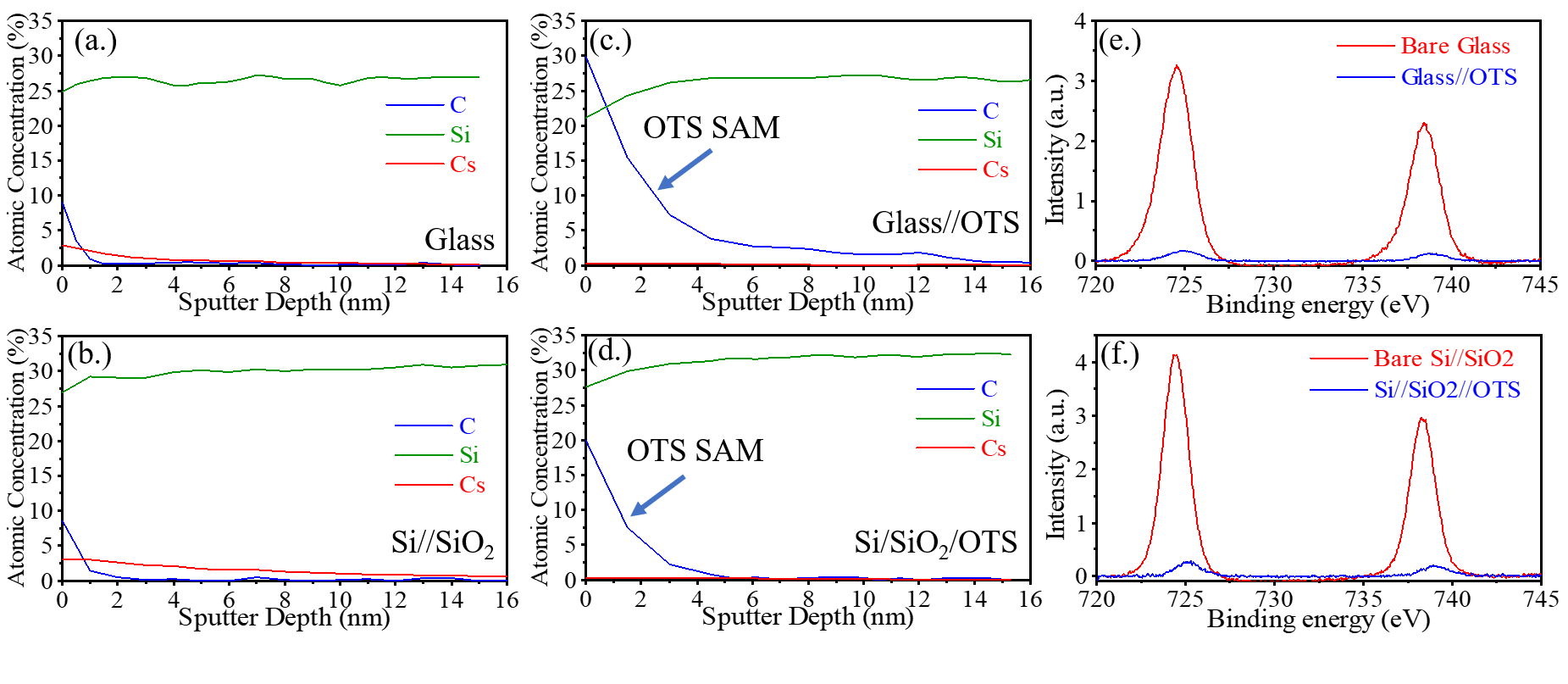}
   \vspace{-30pt}
   \caption{(a) XPS depth profile from a bare glass sample exposed to Cs. The C peak visible on the sample surface, upper $\sim 1\,$nm is from adventitious C on the surface. A distinct non-zero Cs signal is also observed at the surface on the same depth scale. (b) XPS depth profile from a bare $150\,$nm thick SiO$_2$ surface coating on Si. A similar Cs peak to the one found on glass is observed. (c) XPS depth profile from an OTS SAM coated glass surface. Here, the C atomic concentration is much larger and is found to increase at the surface where the Si peak is decreasing. The feature indicates the formation of the SAM. The depth resolution is not high enough to completely separate the signals from the substrate and OTS SAM. The OTS SAM surface is around $2\,$nm thick. A much smaller amount of Cs is found on the surface, see (e). (d) XPS depth profile of an OTS SAM coated SiO$_2$ surface. Again the shape of the C peak and the Si peak shows the OTS SAM formation. A much smaller amount of Cs is detected on the OTS SAM coated surface. Similar to (c), the OTS SAM is around $2\,$nm thick. (e) XPS depth profile showing a comparison of Cs on the glass and OTS SAM coated glass surface, see text for fractions. (f) XPS depth profile showing a comparison of Cs on the SiO$_2$ surface and OTS SAM coated SiO$_2$ surface, see text for fractions.}    \label{fig:XPS}
\end{figure*}

FIG.~\ref{fig:spectra}b shows a room temperature, saturated absorption spectra obtained for the Cs $6$S$_{1/2}(\textrm{F} = 4)\rightarrow 6$P$_{3/2} (\textrm{F}^\prime = 3,4,5)$ transitions using one of the OTS coated vapor cells. The spectra show well-resolved peaks and no indication of collisional broadening. A distinguishing feature of the observations is that a field to repump the atoms out of the $6$S$_{1/2}(\textrm{F} = 3)$ state is not required. OTS coatings are known to preserve nuclear spin polarization when the atoms collide with the walls of the vapor cell for thousands of collisions \cite{BudkerRomalis2007,Straessle2014}. Typically, OTS coatings are applied to large, glass blown vapor cells with sizes on the order of several cm. In these cases, the atoms can be pumped into the $6$S$_{1/2}(\textrm{F} = 3)$ state and no saturated absorption can be observed unless the atoms are repumped. Normally, in an uncoated vapor cell, wall collisions maintain an equilibrium of $6$S$_{1/2}(\textrm{F} = 3)$ and $6$S$_{1/2}(\textrm{F} = 4)$ atoms. Since the time between wall collisions of the atoms is much less in the OTS coated MEMs vapor cell, a repumping laser is unnecessary for observing the saturated absorption spectra.

To characterize the spectral properties in the OTS coated MEMs vapor cells, we performed spectroscopy using the near-Doppler free, three-photon method, where the $6\textrm{S}_{1/2} \rightarrow 6\textrm{P}_{1/2} \rightarrow 9\textrm{S}_{1/2} \rightarrow 42\textrm{P}_{3/2}$ EIA system is utilized \cite{Shaffer2018,Bohaichuk2023}. FIG.~\ref{fig:spectra}c shows a comparison between three different OTS coated MEMs vapor cells and a $2.5\,$cm diameter, $7.5\,$cm long Pyrex vapor cell. The average EIA spectral linewidth is $336\,$kHz with a line shift of $32\,$kHz, relative to the data obtained in the Pyrex vapor cell. The EIA spectral linewidth observed in the Pyrex vapor cell is $290\,$kHz. The average shift corresponds to a field of $\sim 10\,$mV$\,$cm$^{-1}$. The polarizability of the $42\textrm{P}_{3/2}$ state is $301.5\,$MHz$\,(\textrm{V}/\textrm{cm})^{-1}$ \cite{ARC2017}. The EIA experiments were done at room temperature. It is difficult to determine from these experiments if the variations in the spectral line shifts between the control and MEMs vapor cells are significant. It is possible that the background electric fields in the vapor cells are shifting between measurements. The EIA spectral linewidths of the OTS coated MEMs vapor cells are generally broader than the Pyrex reference vapor cell, possibly indicating that background gas collisions are now playing a role. 

FIG.~\ref{fig:XPS} shows x-ray photoelectron spectroscopy (XPS) data from bare and OTS coated silicon and glass samples exposed to a Cs atmosphere. In FIG.~\ref{fig:XPS}a-b, the bare surfaces are shown. A carbon signal corresponding to organic molecules on the surface is detected along with Cs. In FIG.~\ref{fig:XPS}c-d, the results for the OTS coated surfaces are shown. The strong Carbon signal at the surface is indicative of the SAM. Most notably, the Cs on the surface has been drastically reduced. FIG.~\ref{fig:XPS}e-f display a comparison of the Cs on the OTS coated and bare surfaces of Si and glass. The XPS experiments indicate that on bare glass the Cs coverage is $2.9\%$, while for bare Si the Cs coverage is $3.0\%$. In contrast, the Cs coverage on OTS coated glass is $0.2\%$, while for OTS coated Si the Cs coverage is $0.3\%$. Since the electric fields are generated by dipoles formed on the surface, thus proportional to the coverage, the XPS data indicates a reduction of the electric fields near the glass surface by $\sim 14.5$ and on the Si surface by $\sim 10$. The corresponding decrease in Stark shifts are $210$ and $100$, respectively. The reduction of the Stark shift by more than $100$ is consistent with the observed shifts in the OTS coated MEMs vapor cells. In prior work on SiO$_2$ vapor cells, we obtained spectral linewidths of $5-15\,$MHz with similar Stark shifts, depending on the vapor cell \cite{Noaman23}. We attribute the slightly worse results for Si to the fact that the Si frame interior surfaces have a greater surface roughness than the glass, despite being chemically polished.

We have shown that an OTS SAM applied to the interior cavity of a MEMs vapor cell can significantly reduce the adsorption of Cs atoms to the surfaces. The reduction in Cs on the surface reduces the electric fields to $<$10$\,$mV$\,$cm$^{-1}$ and enables the observation of spectral lines on the order of $300\,$kHz in a MEMs vapor cell. Complementing the spectral evidence, we carried out XPS depth profiling studies that showed over an order of magnitude less Cs on the OTS SAM than on either bare glass or Si. The OTS SAM coating was enabled by our low temperature bonding method. Future work involves manufacturing these OTS coated MEMs vapor cells on the wafer scale. The work significantly advances MEMs vapor cells for applications in quantum sensing like magnetometery, electrometry, rotational sensing, and chip scale atomic clocks. 

\begin{acknowledgements}
We acknowledge funding from the Defense Advanced Research Projects Agency under HR001120S006 (SAVaNT) and HR00112530094 (EQSTRA) and The National Research Council Internet of Things: Quantum Sensors Challenge program through Contracts No. QSP-058-1 and No. QSP-105-1 for support.
\end{acknowledgements}

\bibliography{OTSpaper}

\end{document}